\documentclass[reprint,prl,aps,superscriptaddress]{revtex4-1}
\usepackage{times}
\usepackage{graphicx}
\usepackage{amsmath, braket, amsfonts}
\usepackage{amssymb}
\usepackage{natbib}
\usepackage{bm, color}
\usepackage{xcolor}
\usepackage{siunitx}
\newcommand{\m}{moir\'e }
\newcommand{\bperp}{B_\perp}
\newcommand{\bpar}{B_\parallel}
\newcommand{\bx}{B_\parallel^x}
\newcommand{\by}{B_\parallel^y}
\newcommand{\rxx}{R_{xx}}
\newcommand{\idc}{I_{\mathrm{dc}}}
\newcommand{\dvdi}{dV/dI}
\newcommand{\iac}{I_{\mathrm{ac}}}
\newcommand{\rxy}{R_{xy}}
\DeclareUnicodeCharacter{03BD}{{$\nu$}}

\makeatletter
\def\maketitle{
\@author@finish
\title@column\titleblock@produce
\suppressfloats[t]}
\makeatother

\begin{document}

\title{Signatures of a ferro-Josephson effect in twisted graphene}

\author{Ruiheng Su}
\affiliation{Quantum Matter Institute, University of British Columbia, Vancouver, British Columbia, V6T 1Z1, Canada}
\affiliation{Department of Physics and Astronomy, University of British Columbia, Vancouver, British Columbia, V6T 1Z1, Canada}

\author{Zhenxiang Gao}
\affiliation{Quantum Matter Institute, University of British Columbia, Vancouver, British Columbia, V6T 1Z1, Canada}
\affiliation{Department of Physics and Astronomy, University of British Columbia, Vancouver, British Columbia, V6T 1Z1, Canada}

\author{Christopher Coleman}
\affiliation{Quantum Matter Institute, University of British Columbia, Vancouver, British Columbia, V6T 1Z1, Canada}
\affiliation{Department of Physics and Astronomy, University of British Columbia, Vancouver, British Columbia, V6T 1Z1, Canada}

\author{Manabendra Kuiri}
\affiliation{Quantum Matter Institute, University of British Columbia, Vancouver, British Columbia, V6T 1Z1, Canada}
\affiliation{Department of Physics and Astronomy, University of British Columbia, Vancouver, British Columbia, V6T 1Z1, Canada}
\affiliation{Department of Physics, Birla Institute of Technology and Science, Pilani, Hyderabad Campus, Telangana 500078, India}

\author{Dacen Waters}
\affiliation{Department of Physics, University of Washington, Seattle, Washington, 98195, USA}
\affiliation{Intelligence Community Postdoctoral Research Fellowship Program, University of Washington, Seattle, Washington, 98195, USA}

\author{Kenji Watanabe}
\affiliation{Research Center for Electronic and Optical Materials, National Institute for Materials Science, 1-1 Namiki, Tsukuba 305-0044, Japan}

\author{Takashi Taniguchi}
\affiliation{Research Center for Materials Nanoarchitectonics, National Institute for Materials Science, 1-1 Namiki, Tsukuba 305-0044, Japan}

\author{Matthew Yankowitz}
\affiliation{Department of Physics, University of Washington, Seattle, Washington, 98195, USA}
\affiliation{Department of Materials Science and Engineering, University of Washington, Seattle, Washington, 98195, USA}

\author{Nemin Wei}
\affiliation{Department of Physics, Yale University, New Haven, Connecticut, USA}

\author{Chunli Huang}
\affiliation{Department of Physics and Astronomy, University of Kentucky, Lexington, Kentucky 40506-0055, USA}

\author{Allan H. MacDonald}
\affiliation{Department of Physics, University of Texas at Austin, Austin, Texas 78712, USA}

\author{Joshua Folk}
\email{jfolk@physics.ubc.ca}
\affiliation{Quantum Matter Institute, University of British Columbia, Vancouver, British Columbia, V6T 1Z1, Canada}
\affiliation{Department of Physics and Astronomy, University of British Columbia, Vancouver, British Columbia, V6T 1Z1, Canada}

\date{\today}

\begin{abstract}
When a spin-polarized current is driven across a magnetic domain wall, the resulting spin-transfer torque may, beyond a critical threshold, set the wall's moments into precession. This precession modulates the Berry curvature experienced by electrons traversing the wall, producing an electromotive force that is topological in nature and proportional to the precession frequency,  mapping precisely onto the DC Josephson effect and leading to the name ferro-Josephson effect.
We report signatures consistent with this effect in a twisted graphene van der Waals heterostructure, where spin and valley textures are linked by exchange, Hund's coupling, and spin-orbit interactions. Tuned to fillings where the isospin degeneracy is spontaneously broken, the samples develop a sharp peak in the longitudinal resistance within a fraction of a millitesla of $\bpar=0$---a peak that disappears as the current is reduced toward zero. In differential resistance the feature resolves into sharp resonances that disperse with $\bpar$ on microtesla and picoampere scales. We argue that these arise from the current-driven precession of spin-domain-wall moments, in competition with the in-plane anisotropy set by a minuscule applied field, and that they establish nonlinear transport as a sensitive probe of isospin domain-wall dynamics at energy scales far below $k_BT$.
\end{abstract}

\maketitle

Twisting two sheets of graphene together is, by now, an established technique to create an electronic system with bands that are both topological---that is, imbued with significant Berry curvature---and relatively flat. This has led to frequent observations of spontaneous orbital magnetization in twisted graphene devices
~\cite{sharpe2019emergent,serlin2020intrinsic,kuiri2022spontaneous,waters2024topological,suMoiredriven2025,liFractionalHighChernInsulator2026}, where valley polarization determines the sign of the orbital magnetization as the two inequivalent valleys $K$ and $K^{\prime}$ carry opposite Berry-curvature-induced moments. These moments are fixed by the 2D geometry to lie out of the plane, coupling strongly to perpendicular magnetic fields, $\bperp$, but only weakly to $\bpar$ \cite{antebi2022plane,bigeard2024magic,mandal2023valley}.

The hallmark of spontaneous orbital magnetization is the anomalous Hall effect: a hysteresis loop in the transverse (Hall) resistance, $\rxy$, as a function of $\bperp$, with sharp jumps when domain polarizations switch direction.
Indeed, the ubiquity of domains in spontaneously magnetized graphene has been clearly visualized by scanning nanoscale magnetometry\cite{tschirhartImagingOrbitalFerromagnetism2021a, grover2022chern,dutta2026reconfigurable}. Jumps in longitudinal resistivity often coincide with the jumps in Hall resistance, as the conduction electrons scatter strongly when they cross from a domain with one polarization to another that is anti-aligned. 

Another possible source of resistivity across domain walls in magnetic systems---predicted nearly 40 years ago~\cite{berger1986possible} but only once reported in the literature~\cite{yang2009universal,yang2010topological}---has its origin in topology rather than scattering. The flow of polarized current across a domain wall creates a spin torque by conservation of angular momentum. Depending on the domain-wall type and the sample geometry~\cite{mcmichael1997head}, this may cause the spins within the domain wall to precess or, equivalently, vortices within the wall to slide transverse to the current flow. The spin texture within a domain wall endows electrons passing through with a Berry curvature; when the spin texture precesses, the Berry curvature flux changes in time, creating an electromotive force analogous to the Lorentz force due to a changing magnetic flux\cite{yang2010topological}. This has been referred to as the ferro-Josephson effect due to its close analogy to the Josephson effect in superconductors. The additional voltage drop associated with the spin-texture precession is simply $-\delta V = (\hbar/e)\,d\phi/dt$, where $\phi$ represents the spin orientation within the wall, just as voltage drop across a Josephson junction is proportional to the precession frequency of the phase of the superconducting order parameter, $(\hbar/2e)\,d\varphi/dt$.

\begin{figure*}[!ht]
    \centering
    \includegraphics[width=\textwidth]{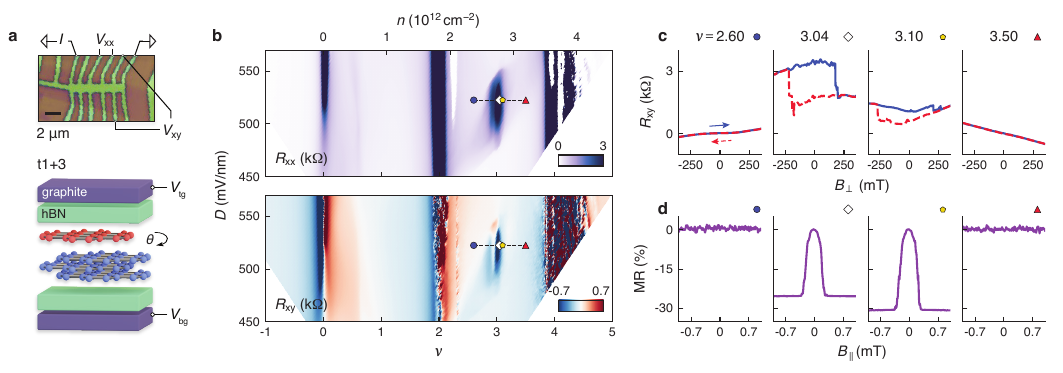}
    \caption{\textbf{Anomalous in-plane magnetoresistance.} \textbf{a}, Optical micrograph of the twisted monolayer-trilayer graphene (t1+3) device labeled with the measurement configuration and schematic rendering of the van der Waals stack with dual graphite gates. \textbf{b}, Longitudinal resistance $R_{\rm{xx}} = V_{\rm{xx}}/I_{\rm{ac}}$ (top) and Hall resistance $R_{xy} = V_{\rm{xy}}/I_{\rm{ac}}$ (bottom) as functions of displacement field $D$ and carrier density $n$ (upper axis) /\m band filling factor $\nu$ (lower axis). Data were field symmetrized (antisymmetrized) at $B_{\perp} = \pm 100$ mT. Colored markers correspond to gate voltage settings ($D = 0.523$ V/nm) for data \textbf{c}, \textbf{d}. \textbf{c}, Out-of-plane magnetic field dependence of $R_{xy}$, showing the anomalous Hall effect near $\nu = 3$. \textbf{d}, In-plane magnetic field dependence of $R_{xx}$ at $B_{\perp} = 0$. The in-plane magnetoresistance, defined as $\mathrm{MR} \equiv (R_{xx}(B_{\parallel}) - R_{xx}(0))/R_{xx}(0)$, reveals an anomalously sharp peak in $R_{xx}$ at $B_{\parallel}=0$, coinciding with the anomalous Hall effect at the same gate voltages.}
    \label{fig:1}
\end{figure*}

Here, we report current-activated transport signatures that are naturally interpreted as ferro-Josephson-like domain-wall dynamics in twisted graphene orbital magnets. When the samples are tuned via gate voltage to regimes of spontaneous orbital magnetization, a sharp increase in longitudinal resistance, $\rxx$, often appears when the in-plane magnetic field is within a few tenths of a millitesla of zero.  Given that in-plane field couples predominantly to electron spin (only weakly to valley or orbital degrees of freedom), this very narrow field scale implies that the sample resistance is strongly affected by Zeeman energies that are 10's of neV or even smaller---a sensitivity to in-plane magnetic field that is, to our knowledge, without precedent in graphene or any other two-dimensional material.   An essential characteristic of this extremely narrow $\rxx(\bpar)$ peak is that it vanishes as the current is reduced toward zero.    We argue that these signatures are likely the result of a ferro-Josephson mechanism.  Spin- and valley-polarized transport current applies a torque to the magnetic moments in the walls between magnetized domains, which lie in the plane of the graphene.  At $\bpar=0$ these moments can easily precess within the plane, leading to excess resistance. But even tens of microtesla of $\bpar$ creates an anisotropy within the plane that blocks precession until the current-induced torque reaches a critical value.

The appearance of a millitesla-wide resistance peak at $\bpar=0$ was, to our knowledge, first reported in twisted double bilayer graphene---two Bernal bilayer twisted atop each other (t2+2)---in Ref.~\cite{kuiri2022spontaneous}. There, it was described as a surprising and unexplained phenomenon that appears alongside the anomalous Hall effect when gate voltages are tuned to values where time reversal symmetry is broken spontaneously.  The $\bpar=0$ peak was unstable in that system, making thorough investigations extremely challenging.  The breakthroughs in understanding presented here were possible due to analogous, but stable, signatures observed in a new sample, a stack of monolayer graphene twisted above a trilayer (t1+3) with a twist angle of $1.29^\circ$ (Fig.~1a, Methods).  Both t2+2 and t1+3 are members of the broader t$M$+$N$ family---twisted stacks of Bernal-stacked $M$- and $N$-layer graphene---which share a common hierarchy of correlated and topological states in the lowest \m conduction band~\cite{waters2024topological}.

Figure 1 provides an overview of the device.
$\rxx$ and $\rxy$ are mapped out in Fig.~\ref{fig:1}b, across the lowest \m conduction band within a narrow range of $D$, over which the conduction band resides within the trilayer component of the sample.  (The complete gate-tuned phase diagram is shown in ED Fig.~\ref{efig:full_map_s1} and in Ref.~\onlinecite{waters2024topological}.)
Insulating states are seen at $\nu=2$ and $3$, indicating spontaneous breaking of the four-fold spin and valley band degeneracy, together called isospin.  We focus on the insulating state at $\nu=3$, where a robust anomalous Hall effect appears over a sliver of space in the $\{\nu,D\}$ plane (Fig.~1c and ED Fig.~\ref{efig:AHE_s2}),  indicating that the system is valley polarized with spontaneously broken time-reversal symmetry.  Further measurements of the $\nu=3$ state in this sample, presented in Ref.~\cite{waters2024topological}, identify this state as an incipient Chern insulator with $C=-2$. Exclusively within the sliver of the $\{\nu,D\}$ plane where the t1+3 state is valley polarized, we observe that $\rxx$ is narrowly peaked at $\bpar=0$ (Fig.~1d, see also ED Fig.~\ref{efig:SHF_s3}), with the resistance dropping sharply above a fraction of a millitesla.
It is important to note that the data in Fig.~\ref{fig:1} were taken with alternating current bias $\iac=1.4$~nA applied, similar to the $\iac=1$~nA bias used in Ref.~\cite{kuiri2022spontaneous}. This value will be seen to be relatively large in the context of the data that follows.  Although it corresponds only to $\mu$V-scale voltages dropped across the region of the sample between voltage probes, it does not represent the zero bias, linear-response limit.

\begin{figure*}[!ht]
    \centering
    \includegraphics[width=\textwidth]{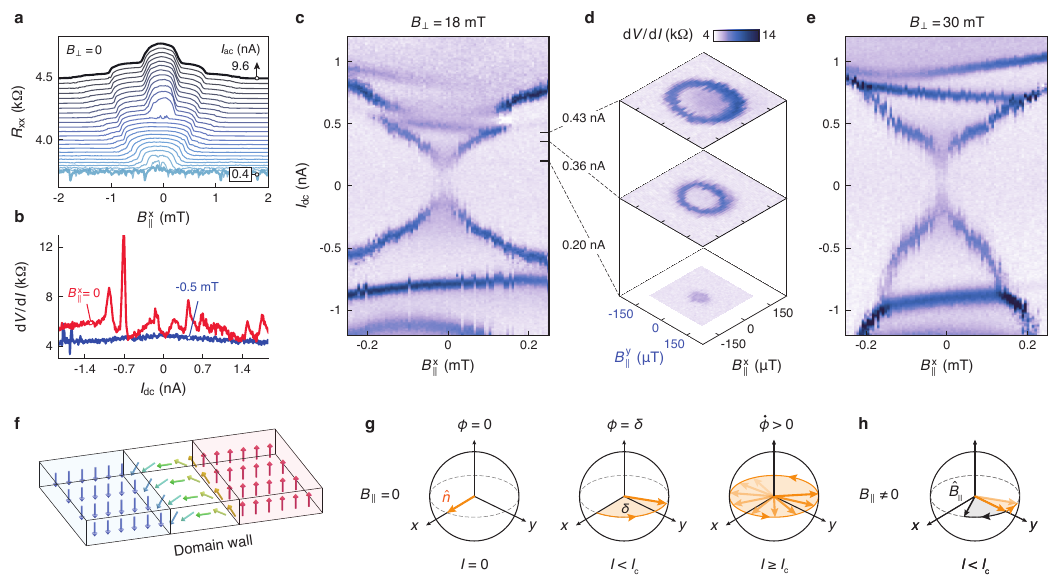}
    \caption{\textbf{Signatures of a ferro-Josephson effect.} 
    \textbf{a} $R_{xx}(B_{\parallel})$ obtained at $\nu = 3.02, D = 0.523$ V/nm, for applied AC bias ranging from $I_{\mathrm{ac}} = 0.4$ to 9.6 nA. Traces are vertically offset by 50~$\Omega$ for clarity; each corresponds to a 0.4 nA increment in $I_{\mathrm{ac}}$. \textbf{b}, Differential resistance obtained at gate voltages as in \textbf{a}, showing nearly no dependence on the DC bias ($\mathit{I}_{\mathrm{dc}}$) at $\mathit{B}_{\parallel}^{x} = -0.5$ mT, and sharp resonances at no applied $\mathit{B}_{\parallel}^{x}$. Data was obtained at $B_{\perp} = 20$ mT. \textbf{c,e}, Differential resistance as a function of DC bias and $B_\parallel^x$, at $B_\perp = 18$ mT (\textbf{c}) and $30$ mT (\textbf{e}). \textbf{d}, Constant-current planar cuts of $dV/dI$ across the $\{B_\parallel^x, B_\parallel^y\}$ plane at $I_{\mathrm{dc}} = 0.20, 0.36$ and $0.43$ nA. \textbf{f}, Schematic rendering of a domain wall between domains of opposite spin polarization. \textbf{g}, Orientation of in-plane spin moment $\hat{n} = (\cos(\phi), \sin(\phi))$ in the domain wall in the absence of an applied $B_{\parallel}$, for zero bias $(I=0)$ and for bias below and above a critical value $(I<I_{c}$, $I\geq I_{c})$. \textbf{h}, A small but finite applied $B_{\parallel}$ competes with the reorientation of $\hat{n}$ by the transport current. 
    }
    \label{fig:2}
\end{figure*}

The central discovery of this paper is summarized in Fig.~2. The width and shape of the $\bpar=0$ peak in $\rxx$ are strongly dependent on $\iac$ (Fig.~2a), disappearing entirely (in this case) for biases below 0.4~nA.  For higher $\iac$, an inner peak appears atop the primary one.  The origin of this behaviour becomes clear when the differential resistance, $dV/dI|_{\idc}$, is measured by locking onto a small AC modulation on top of a variable DC bias, $\idc$ (Fig.~\ref{fig:2}b), then $\dvdi$ traces are measured across the $\bpar$ magnetoresistance peak.  Figure~\ref{fig:2}c shows an example of such a measurement, in which it becomes clear that the elevated $\rxx$ at $\bpar=0$ seen in Figs.~\ref{fig:1}d and \ref{fig:2}a, collected with $\iac$ on the nA scale,  are in fact the accumulated effect of multiple peaks in $\dvdi$.  At $\bpar=0$, these peaks approach zero bias, but with increasing $\bpar$ the first peak in $\dvdi$ moves to higher and higher bias, and above some field disappears entirely (see ED Fig.~\ref{efig:suppresion_s4}).

The $\dvdi$ spectra were observed to depend roughly--though not exactly--isotropically on magnetic field direction within the plane, but to be barely affected by out-of-plane  magnetic fields that are orders of magnitude larger.  The detailed spectrum in Fig. 2c reflects $\dvdi$ as $B_\parallel^x$ is scanned through zero, but symmetry in the $\{\bx,\by\}$ plane is clear from the ring-shaped patterns that emerge when data like that in Fig. 2c is collected at fixed $\idc$ across the 2D $\{B_x,B_y\}$ plane (Fig.~2d).  In contrast, the $\bpar$ fingerprints change only weakly when $\bperp$ is changed from 18~mT (Fig.~\ref{fig:2}c) to 30~mT (Fig. 2e).

The data in Figs.~\ref{fig:1} and \ref{fig:2} highlight four essential characteristics against which any possible microscopic explanation must be tested. First, the phenomenon appears only when valley degeneracy is broken, that is, where AHE also appears.  Second, the higher-resistance state requires a non-zero current to activate.  That activation current may be vanishingly small when $\bpar$ is exactly zero, but it grows with $\bpar$ in a way that is approximately linear for small $I_{dc}$.  Third, the phenomenon is extremely anisotropic comparing in-plane to out-of-plane directions, though nearly isotropic within the plane.  Fourth, the energy scales (Zeeman energy and voltage bias) associated with this effect are much smaller than $k_B T$, suggesting that conventional quasiparticle scattering mechanisms are not responsible. These characteristics were confirmed also to describe the $\bpar$ magnetoresistance peak previously observed in t2+2 (see supplement).  Taken together, they leave few plausible mechanisms that could be responsible for this effect.

\begin{figure*}[!ht]
    \centering\includegraphics[width=\textwidth]{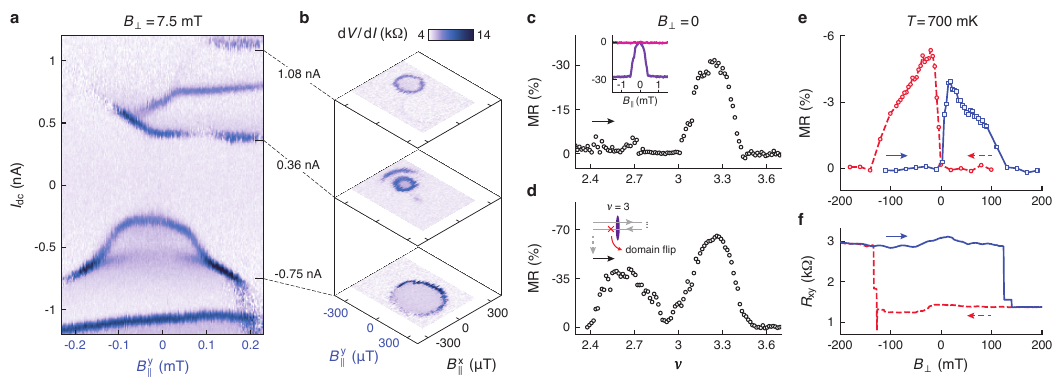}
\caption{\textbf{Role of $\bperp$ and domain preparation in stabilizing the ferro-Josephson effect} \textbf{a}, $dV/dI$ map versus $\by$ and $\idc$ at $B_{\perp} = 7.5$ mT. \textbf{b}, In-plane magnetic field dependence of $dV/dI$ at fixed $\idc = 1.08, 0.36$ and -0.75 nA, showing a shift in the central feature between positive and negative current bias. \textbf{c, d}, Dependence of the in-plane magnetoresistance (here calculated at $\bpar= 1$ mT) on the history of the gate voltage. MR sequentially measured for increasing $\nu$, starting from $\nu = 2 + 0.27$ (indicated by the right pointing arrow), and along a contour of $D = 0.523$~V/nm ($B_{\perp} = 0$). Inset: representative MR traces at $\nu = 2.8$ and $\nu = 3.2$. \textbf{d}, MR measured after a sequence of gate voltage sweeps between $\nu = 3.02 \pm 0.11$ prepares a resistively detected domain flip. \textbf{e, f}, Dependence of MR and $R_{xy}$ hysteresis on $B_{\perp}$, measured at $\nu = 3.02, D = 0.523$ V/nm, $T = 700$ mK. Arrows indicate the direction that $B_{\perp}$ was swept. Following initialization into a fully polarized state by sweeping $B_{\perp}$ past the coercive field $B_{c}$ of the anomalous Hall effect, pronounced MR appears as $\vert B_{\perp} \vert$ is reduced through $0$ from $\vert B_{\perp}\vert > \vert B_{c} \vert$. 
$I_{\mathrm{ac}} = 1.4$ nA was used in the MR measurement \textbf{c}-\textbf{e}.}
    \label{fig:3}
\end{figure*}

Before turning to a microscopic mechanism, we first exclude the most mundane alternative: ordinary current-driven depinning, in which spin-transfer torque slides a domain wall past disorder pinning sites above a threshold current. Two features argue strongly against it. First, the activation current tracks the in-plane field: it grows approximately linearly with $\bpar$ and collapses toward zero as $\bpar\to0$ (Fig.~\ref{fig:2}), down to a field scale equivalent to an anisotropic in-plane Zeeman energy of order $10$~neV. In contrast, a conventional depinning threshold would be determined by the pinning potential rather than $\bpar$. The same reasoning excludes Joule heating, whose onset would not track $\bpar$. 
Second, depinning translates the domain wall irreversibly, in hysteretic jumps, whereas the signal we study is reproducible and reversible, growing smoothly with bias and recurring on every sweep.

The sole mechanism we have identified that is consistent with all observations is the ferro-Josephson effect~\cite{berger1986possible}. To connect the ferro-Josephson effect with our observations, we add two known characteristics of graphene-based van der Waals systems: first,  robust spin and valley domains frequently appear in these systems, apparently stabilized by disorder ~\cite{grover2022chern, tschirhartImagingOrbitalFerromagnetism2021a}.  Second,  graphene hosts a Kane-Mele-type spin-orbit interaction that, while weak compared to the strength of spin-orbit in many other materials, is understood to be many 10's of $\mu$eV, equivalent to hundreds of millikelvin, and dominates the spin physics of graphene at low field \cite{kane2005quantum,PhysRevLett.122.046403,banszerus2020observation,kurzmann2021kondo}.  This interaction locks the spin of electrons in graphene to the valley degree of freedom at low energy scales.  (An additional spin-valley locking interaction of similar magnitude has been predicted in ABA-stacked trilayer graphene \cite{mccann2010spin}.)

With these ingredients, we present a minimal calculation of a ferro-Josephson mechanism that explains these observations. We consider current flow between two adjacent magnetic domains of opposite spin polarization (Fig.~\ref{fig:2}f), across which the spin polarization is expected to rotate smoothly between $\ket{\downarrow_z}$ and $\ket{\uparrow_z}$ due to a Hund's coupling that is significantly larger than the spin-orbit coupling (see companion theoretical study \cite{das2026multicomponent} and Methods).
Flowing a current, $I$, between them implies a rate of change of the z-component of angular momentum, imposing a spin-transfer torque $\tau_{I}=\hbar I/e$, on the spin moments within the domain wall.  If the system is to be in steady-state, this must be balanced by a restoring torque, $dU/d\phi$, that reflects the energy cost to rotate the domain wall moments within the plane. The gradual rotation of spin moments across the domain wall implies that in the middle of the wall the moments will have rotated into the plane (see methods), at which point the contribution of an in-plane field to the domain wall energy will include a term $N_\parallel \mu_B\bpar \cos(\phi)$, where $N_\parallel$ is the number of spin-1/2 moments within the wall that are rotated into the plane and the g-factor is assumed to be 2.  The result is a restoring torque $dU/d\phi= N_\parallel \mu_B\bpar \sin(\phi)$.

With this framework in mind, let us consider the $I-V$ characteristics at non-zero $\bpar$.  Near zero current, $\dvdi$ is just set by conventional transport-limiting scattering channels, such as momentum scattering or scattering at domain wall boundaries.  The in-plane domain wall moments align with the albeit-tiny external field because no other term breaks the in-plane isotropy  (Fig.~\ref{fig:2}h). As $I$ increases, the domain wall moments twist away from $\bpar$ such that the spin-transfer torque matches the restoring torque, $\hbar I/e=2N_\parallel \mu \bpar \sin(\phi)$, until, for large enough $I$, $\phi$ reaches $\pi/2$ and this equation can no longer be satisfied---a threshold current $I_c=N_\parallel\mu_B\bpar\,e/\hbar$ that is proportional to $\bpar$ and vanishes as $\bpar\to0$, so that the $\dvdi$ peak collapses to zero bias in that limit.  At that point, no steady-state solution is possible and the domain wall moments will precess (Fig.\ref{fig:2}g), leading to the Josephson-like voltage drop whose value depends on how strongly the precession is damped--a parameter related to Gilbert damping in ferromagnets.  The onset of precession produces a peak in $\dvdi$ ($V$ increases rapidly with $I$), but once the precessing domain wall moments are again in steady-state due to a balance between damping and torque, $\dvdi$ will drop to the value set by conventional transport scattering channels.

This predicted linear-in-$\bpar$ threshold can be compared directly to experiment: where the resonance disperses cleanly its slope, $dI_c/d\bpar$, is a few ~nA/mT (e.g. Fig.~\ref{fig:2}c,e and ED Fig.~\ref{efig:full_dvdi}), corresponding via $N_\parallel\mu_B\,e/\hbar$ to a wall of $N_\parallel\sim 200$ precessing moments.  The slope varies between datasets and even between in-plane field directions, but stays of this order wherever a clear linear dispersion appears.

The mechanism described above is also consistent with the resilience of the effect to moderate values of $\bperp$ (see methods), as the domains already have moments locked out-of-plane due to the exchange interaction.  Small amounts of $\bperp$ will simply shift the relative energy of the two out-of-plane polarizations, leaving the domain wall structure approximately intact and preserving the ability for free precession in the plane.  In fact, a small amount of $\bperp$ was important for stabilizing the resistivity signatures enough that a full 2D map could be obtained, and for generating the relatively symmetric-in-$\bpar$ data shown in Fig.~\ref{fig:2}.  At $|\bperp|\lesssim 5$~mT, the data in 2D maps analogous to  those shown in Fig.~\ref{fig:2} would often jump in the middle of the scan, to a domain configuration where no $dV/dI$ features were visible (see ED Fig.~\ref{efig:full_dvdi}).

Figures~\ref{fig:3}a and \ref{fig:3}b illustrate the low-$\bperp$ behaviour in detail.  A slice taken at $\bperp=7.5$~mT is significantly less symmetric (Fig.~\ref{fig:3}a), compared to those taken at $\bperp=18$ or 30~mT (Fig.~\ref{fig:2}c,e).  The lack of symmetry is especially obvious in the constant-current slices (Fig.~\ref{fig:3}b), with no single symmetry point visible in the data.  As well, a larger $\bpar\rightarrow 0$ residual appears in the threshold current, compared to what was seen in Fig.~\ref{fig:2}.  We speculate that this finite bias required to induce dissipation may result from domain geometry combined with spin--orbit coupling.  The threshold current offset, like the asymmetry, grows as $\bperp\to0$, perhaps because the domain polarizations are less firmly locked out of plane in that limit.

Clear evidence for the domain nature of this effect may be found in how the strength of the $\bpar=0$ feature depends on the recent history of the sample.  The $\bpar$-dependent features that are the subject of this work were not observed in every scan, even at gate settings where they usually appear.  This irreproducibility may be explained by differing domain structures that depend on sample history, recalling that the ferro-Josephson signatures described above require current to flow across a domain wall between domains with different spin polarization.    Figures~\ref{fig:3}c--f explore two approaches for resetting the domain polarizations in the sample, first using gate-voltage sweeps into the valley-polarized state near $\nu=3$, then using $\bperp$-sweeps to generate analogous effects.  For both, $I_{\rm ac}=1.4$ nA measurements analogous to those presented in Fig.~\ref{fig:1}d are used to characterize the strength of the in-plane magnetoresistance (MR), without the complexity of $\dvdi$ spectra like those in Fig.~\ref{fig:2}.

Figure~\ref{fig:3}c shows the MR as gate voltage is stepped along a contour of constant $D=0.523$~V/nm across the $\nu=3$ state, starting from $\nu=2.27$, where the sample is not valley-polarized (see ED Figs.~\ref{efig:AHE_s2} and \ref{efig:SHF_s3}).  The MR remains zero until the $\nu=3$ state is crossed.  This was typically repeatable on scanning once back across $\nu=3$; but after several passes back and forth within the valley-polarized region ($2.8<\nu<3.2$), the sample underwent a discrete, resistively detected domain flip, after which large MR appeared on both sides of $\nu=3$ (Fig.~\ref{fig:3}d).  Such flips are abrupt, one-time reconfigurations of the domain pattern---readily distinguished from the smooth, reproducible current-driven resonances of Fig.~\ref{fig:2}---and their occurrence is direct evidence that the MR tracks the domain configuration.  While we do not have a complete picture of how the domain polarizations are set during these sweeps, Figs.~\ref{fig:3}c,d indicate that domain alignment in the valley-polarized region at low $\bperp$ is strongly affected by gate sweeps across $\nu=3$.

Training by an out-of-plane magnetic field provides an alternative---and less random---route to modifying the domain configuration. It was observed, in t1+3 as in t2+2, that raising $\bperp$ past the AHE coercive field and then returning toward zero eliminates the in-plane MR.  We attribute this to an alignment of the participating domains above $B_c$.  For $\nu$ very close to 3 in t1+3, however, we found that MR would re-emerge as soon as $\bperp$ was scanned through zero (Figs.~\ref{fig:3}e, \ref{fig:3}f), apparently causing some domains to invert and recovering the situation where current flow involves carriers moving between oppositely polarized domains leading to spin torque on the domain wall moments.  The energetics of this process are described further in Methods.

Figure~\ref{fig:3}e illustrates the lack of MR after ramping $\bperp$ past $B_c\approx -130$~mT (Fig.~\ref{fig:3}f), then back towards zero (blue data), but also that the MR immediately reappears upon crossing $\bperp=0$.    If the steps in $\bperp$ were immediately reversed, bringing $\bperp$ back to and past zero, this MR behaviour was repeatable: present for $\bperp>0$, absent for $\bperp<0$.  If, however, the scan was continued past $B_c=+130$~mT then back toward zero, the MR was gone on the positive side of zero but reappeared after crossing to the negative side (red data).  Care was taken to sweep $\bpar$ through the real zero in order to collect these MR data, compensating for any misalignment in the sample, with further details in ED Fig.~\ref{efig:mag_calib}.

The domain-preparation experiments of Fig.~\ref{fig:3} point to a possible domain structure behind the effect. The reappearance of the in-plane MR the instant $\bperp$ crosses zero (Fig.~\ref{fig:3}e,f)---at fields two orders of magnitude below the AHE coercive field---cannot arise from reversing a valley-polarized orbital magnet. It requires, instead, a domain whose spin flips with essentially no coercive field.

This leads to a picture in which a valley-polarized $\nu=3$ domain---the orbital magnet responsible for the AHE, held out of plane by spin-orbit locking---sits beside a low-anisotropy region of negligible coercive field, plausibly the spin-polarized, valley-unpolarized half-metal that generically fills the band between $\nu=2$ and $\nu=4$ in t$M$+$N$ samples~\cite{waters2024topological}. Training in $+\bperp$ aligns the spins of both. On reversal to a small $-\bperp$, the low-anisotropy region flips at once while spin-orbit locking holds the valley-polarized domain, leaving two counter-polarized spin domains separated by a wall---precisely the configuration the ferro-Josephson mechanism requires. Only one of the two domains need be valley-polarized: the mechanism draws only on the difference in spin polarization across the wall, which survives whatever the valley character of the other.

This may distinguish t1+3 from the earlier t2+2 measurements, where the instability appeared tied more directly to the discrete switching of valley-polarized orbital domains, with finite coercive field~\cite{kuiri2022spontaneous,zhenxiangnew}. The stability of the effect in t1+3, and its reappearance on crossing $\bperp=0$, requires the coercive-field-free domain identified here. Although the half-metal state provides a natural explanation for the low-anisotropy component observed in t1+3, there may be others; the phenomenological model for the ferro-Josephson effect is insensitive to the details.

Finally, we note that the data in Fig.~\ref{fig:3}e were  collected at $T=700$~mK, where $k_BT$ exceeds the tens-of-neV Zeeman scale by more than three orders of magnitude. The robustness of the MR that is the central phenomenology of this experiment is particularly telling, and clearly not consistent with thermal or voltage activation over an energy barrier. It is, however, consistent with the ferro-Josephson-like dynamics described above: a current-driven excitation of low energy collective modes, within domains held together by much larger exchange and spin--orbit energies, where the role of the small in-plane field is  merely to set the threshold current.

We caution, however, that Berger's description in Ref.~\cite{berger1986possible} applies in the adiabatic limit where the carrier spin follows the
local exchange field, and may therefore not be directly applicable to the graphene system.  Theoretical estimates for domain-wall widths in t$M$+$N$ orbital ferromagnets  indicate a few moir\'e wavelengths in twisted bilayer graphene Chern insulators~\cite{kwan2021domain}, and two to three times the interparticle spacing in the quarter-metal of rhombohedral graphene~\cite{das2026multicomponent}.  Both estimates yield length scales around 50~nm, not much larger than the interparticle spacing or Fermi wavelength in t1+3 and therefore not deeply in the adiabatic limit.  We speculate that, for finite-length domain walls, the Berger picture may be thought of as the semi-classical description of an inelastic tunneling process across domain walls involving the emission of collective modes into the wall.
The extrapolation of the Berger mechanism down to the tunneling limit is an important direction for future study.

From an experimental point of view, the proposed ferro-Josephson interpretation could be tested by superposing a GHz drive on the DC bias to produce Shapiro-step-like features that would directly confirm the Josephson relation $-\delta V=(\hbar/e)\,d\phi/dt$. Once confirmed, this effect itself would then become a probe sensitive to spin and orbital textures inaccessible by other techniques. As isospin-ordered states appear generically across the t$M$+$N$ family and beyond, this diagnostic would be broadly applicable in twisted graphene and other two-dimensional orbital magnets.

\section{Methods}
\textbf{Device fabrication.} The t1+3 device studied here was previously reported in Ref.~\onlinecite{waters2024topological}, where fabrication details are given. Briefly, graphene flakes containing monolayer and trilayer regions were exfoliated, and the monolayer portions were separated from the trilayer using local anodic oxidation nanolithography~\cite{Li2018,Chen2019a,Saito2020}. The vdW heterostructure was assembled from the top down, such that the monolayer and trilayer flakes---with a relative twist---were encapsulated between hBN dielectrics and top and bottom graphite gates. The completed stack was released onto a Si/SiO$_2$ wafer and etched into a Hall bar geometry, with Cr/Au edge contacts~\cite{wang2013one}.

\textbf{Transport measurements.} The measurements were performed in a Bluefors LD dilution refrigerator equipped with a 3-axis superconducting vector magnet. The nominal base mixing chamber temperature was $T = 10$~mK, as measured by a factory-supplied RuO$_x$ sensor. Four-terminal lock-in measurements were performed by sourcing a small alternating current of $\iac < 1.5$~nA at a frequency $<42$~Hz. Two types of measurements are reported in this paper: ``standard'' lock-in measurements of $\rxx$ and $\rxy$ with $\iac=1.4$~nA, and differential resistance ($\dvdi$) measurements in which $\iac$ was reduced to between $10$ and $80$~pA and added to a direct-current bias through a resistive divider. When specified, $R_{xx}$ and $R_{xy}$ were symmetrized or antisymmetrized according to $R_{xx} = [R_{xx}(B) + R_{xx}(-B)]/2$ and $R_{xy} = [R_{xy}(B) -R_{xy}(-B)]/2$. To reduce contact resistance, a global bottom-gate voltage between $3.0$ and $55.8$~V was applied to the doped Si substrate.

The carrier density, $n$, and the out-of-plane electric displacement field, $D$, were defined as $n= \left(C_{\text{b}} V_{\text{b}}+C_{\text{t}} V_{\text{t}}\right) / e$ and $D=\left(C_{\text{t}} V_{\text{t}} - C_{\text{b}} V_{\text{b}}\right) / 2 \epsilon_0$, where $C_{\text{t}}$ and $C_{\text{b}}$ are the top and bottom gate capacitances per unit area, $e$ is the elementary charge, and $\epsilon_0$ is the vacuum permittivity. The gate capacitances can be estimated using the slope of the gate-voltage dependence of the Hall density, $n_{H} = 1/(e R_{H})$, with $R_{H} = R_{xy}/B_{\perp}$ the antisymmetrized Hall coefficient. The \m band filling factor $\nu$ was defined as $\nu = n/(n_{s}/4)$, where $n_{s}$ is the superlattice density required to fill a four-fold degenerate \m band, given by $n_s=4\frac{2\theta^2}{\sqrt{3}a^2}$, where $a=0.246$~nm is the graphene lattice constant. The Hall filling, used in ED Fig.~\ref{efig:SHF_s3}, was defined analogously as $\nu_{H} = n_{H}/n_{s}$.

\textbf{Phenomenological model.} In this section, we propose a simple phenomenological model to interpret the $\bpar$-dependent nonlinear $I$-$V$ characteristics observed around the $\nu=3$ state where an anomalous Hall effect appears. We focus on the small-but-finite $B_{\perp}$ regime, where two adjacent domains have opposite spin polarizations and the spin continuously rotates   from $\uparrow$ to $\downarrow$ across the domain wall. A companion microscopic study Ref.~\cite{das2026multicomponent} shows that a domain wall connecting the spin- and valley-polarized states $\ket{K\uparrow}$ and $\ket{K'\downarrow}$, 
can reconstruct into a four-component spin-valley texture when intervalley Hund's coupling dominates spin-orbit coupling. Because the reconstructed wall admixes $\ket{K\downarrow}$ and $ \ket{K'\uparrow}$, it has finite in-plane spin moment. 
    
Let the total in-plane moment of the wall be,
\begin{equation}
    \mathbf{M}_{\parallel}
    =
    M_{\parallel}
    \left(
    \cos\phi,\,
    \sin\phi
    \right),
\end{equation}
where $\phi$ is its azimuthal orientation. For a straight wall of length $L_y$, the magnitude of this moment is
\begin{equation}
    M_{\parallel}
    =
    \frac{g\mu_B}{2}
    L_y
    \int dx\,
    \left|n_e(x)\right|
    s_{\parallel}(x),
    \label{eq:wall_moment}
\end{equation}
where $n_e(x)$ is the local carrier density and
\begin{equation}
    s_{\parallel}(x)
    =
    \left|
    \left\langle
    \mathbf{s}_{\parallel}(x)
    \right\rangle
    \right|
\end{equation}
is the transverse spin polarization of the microscopic domain wall profile, which is peaked inside the domain wall, see Ref.~\cite{das2026multicomponent}. An in-plane magnetic field with orientation $\phi_B$ produces the Zeeman energy
\begin{equation}
    U_B(\phi)
    =
    -M_{\parallel}B_{\parallel}
    \cos\left(\phi-\phi_B\right).
    \label{eq:wall_zeeman}
\end{equation}
We also allow for a weak intrinsic in-plane anisotropy,
\begin{equation}
    U_{\rm{ani}}(\phi)
    =
    -K_\perp
    \sin^2\left(\phi-\phi_0\right),
\end{equation}
so that the total $\phi$-dependent domain wall energy is
\begin{equation}
    U(\phi)
    = U_B(\phi) + U_{\rm{ani}}(\phi).
\end{equation}

A current $I$ flowing between oppositely spin-polarized domains transfers angular momentum to the wall. The rate at which an electron crosses the wall is
$\frac{dN}{dt}= \frac{I}{e}$. Because the spin angular momentum changes by $\hbar$ when a fully polarized carrier passes from one domain to the other, the spin-transfer torque is
\begin{equation}
    \tau_I
    =\frac{\hbar I}{e}.
    \label{eq:current_torque}
\end{equation}
Assuming that the wall position is pinned and that its internal dynamics are overdamped, the equation of motion for a spatially uniform collective coordinate $\phi$ is described by 
\begin{equation}
    \Gamma\dot{\phi}
    +\frac{\partial U}{\partial\phi}
    = \frac{\hbar I}{e},
    \label{eq:phase_dynamics}
\end{equation}
where $\Gamma$ is the effective damping coefficient. To the extent that electron traversal of the domain wall is non-adiabatic, the right hand side would be amended to $P\frac{\hbar I}{e}$ where $P<1$ accounts for non-adiabaticity.

A static solution $\dot \phi=0$ exists only when the current-induced torque does not exceed the maximum restoring torque,
\begin{equation}
    \frac{\hbar |I|}{e}\leq\max_{\phi}
    \left|\frac{\partial U}{\partial\phi}\right|.
\end{equation}
When the Zeeman energy dominates the intrinsic in-plane anisotropy, the maximum restoring torque is
\begin{equation}
    \max_{\phi}\left|\frac{\partial U_B}{\partial\phi}\right|
    =M_{\parallel}|B_{\parallel}|.
\end{equation}
Therefore, the critical current increases linearly with applied field:
\begin{equation}
    I_c(B_{\parallel})
    =\frac{eM_{\parallel}}{\hbar}|B_{\parallel}|.
    \label{eq:linear_threshold}
\end{equation}

When $I>I_c$, no static wall orientation can balance the current-induced torque, and the $\phi$ must precess. Using  Eq.~\eqref{eq:linear_threshold}, the equation of motion can be written as
\begin{equation}
\dot{\phi}=\frac{\hbar}{e\Gamma}\left(I-I_c\sin\phi\right),
\end{equation}
The time required for one complete $2\pi$ rotation is
\begin{align}
T=\frac{e\Gamma}{\hbar}\int_0^{2\pi}\frac{d\phi}{I-I_c\sin\phi}\nonumber=\frac{2\pi\Gamma e}{\hbar\sqrt{I^2-I_c^2}}.
\end{align}
The average phase velocity is $\overline{\dot{\phi}}=\frac{2\pi}{T}$ and this give rise to the time-averaged ferro-Josephson voltage
\begin{equation}
    \overline{V}_{\mathrm{FJ}}
    = \frac{\hbar}{e}\frac{2\pi}{T}=
    R_{\mathrm{dw}}
    \sqrt{I^2-I_c^2}\,
    \Theta\left(|I|-I_c\right),
    \label{eq:average_voltage}
\end{equation}
where $R_{\mathrm{dw}}= \frac{\hbar^2}{e^2\Gamma}$.
The measured voltage contains both the ordinary transport background and the dynamical domain-wall contribution,
\begin{equation}
    V=R_0 I+ \overline{V}_{\rm{FJ}}.
\end{equation}
Thus, the square-root onset of the $V_{\rm FJ}$ at the threshold current produces a peak in the differential resistance.

We now provide an estimate of how even a weak in-plane magnetic field can produce a measurable shift in the peak of $dV/dI$.  Let $N_{\parallel}\equiv\frac{2M_{\parallel}}{g\mu_B}$  be the  number of fully polarized spin-1/2 moments in the wall. At threshold, the current-induced torque balances the maximum Zeeman restoring torque so the $I$-$V$ peak shifts with the field strength as follows:
\begin{equation}
\frac{dI_c}{d|B_{\parallel}|}=\frac{e g\mu_B}{2\hbar}N_{\parallel}.
\end{equation}

For an order-of-magnitude estimate, suppose that the carrier density $n_e$ and transverse $M_\parallel$ polarization are approximately uniform over a domain wall of width $\lambda$ and length $L_y$,
then $N_{\parallel}=
\overline{s}_{\parallel}
n_e\lambda L_y$. Here $n_e
=\frac{\nu}{A_M}$ where $A_M=\frac{\sqrt{3}}{2}a_M^2$ is the moire unit cell area. Using $\lambda\sim5a_M$, $L_y\sim10a_M$, $\nu=3$ and $ \overline{s}_{\parallel}\sim 1$, we obtain $N_{\parallel}\sim 2\times10^2$.  Using $g=2$, we find $dI_c/dB_{\parallel}\sim 2.5$ \si{\nano\ampere\per\milli\tesla},
which agrees within a factor of two to the field-dependence of the $I$-$V$ peak dispersions observed throughout this experiment. Although these values of $\lambda$ and $L_y$ are hypothetical, they are not unreasonable given the visibility of domain-dominated behaviour in a micron-sized sample, and are consistent with the spatial extent of analogous domains previously observed by scanning microscopy in similar graphene systems~\cite{tschirhartImagingOrbitalFerromagnetism2021a,grover2022chern}.

A few comments are in order:

Firstly, we note that the appearance of multiple $dV/dI$ peaks in data like Figs.~\ref{fig:2}c,e may be related to ferro-Josephson precession in multiple domain walls within the part of the sample between the voltage probes. If the current crosses these domain walls in parallel (across the sample width), then the appropriate data to compare with a predicted $I_c(\bpar)$ would be a fraction of the applied $\idc$. 

Secondly, since the magnetic system breaks the time-reversal symmetry, the local resistivity and the current distribution in the sample are allowed to be different when switching the direction of the current. As a result, the critical $\idc$ for the nonlinear $I-V$ features is also expected to be asymmetric about the direction of the current.

Thirdly, our model applies only to the small $B_{\perp}$ regime. In this regime, increasing the field in the $-B$ direction reduces the domain wall width $\lambda$ and the in-plane spin polarization $\propto N_{\parallel}$. This, in turn, enhances the resistance $R$ above the critical current. Therefore, our model can explain why the magnetoresistance grows with $|B_{\perp}|$ in the small $B_{\perp}$ regime. However, the suppression of MR signals at large $B_{\perp}$ is not captured in our model. For instance, when $g{\mu_B}B_{\perp} + \Delta_I\sim 0$, the spin in the high-anisotropy domain might deviate from $z$ direction, leading to larger in-plane spin polarization and smaller $R$. We leave the analysis of the large $B_{\perp}$ regime for future study.

The magnetic anisotropy energy of a valley-unpolarized half-metal arises from the competition between the (ferromagnetic) Hund's coupling and spin--orbit coupling. Since the former is typically stronger than the latter in pristine multilayers, the resulting ferromagnet is an XY ferromagnet, with spins polarized on the equator of the spin Bloch sphere. Consequently, in the absence of an in-plane magnetic field $B_{\parallel}$, the spin-domain structure is likely a continuous rotation from out-of-plane (up) to an in-plane (x) direction. 

\textbf{On the emergence of low- and high-anisotropy domains in t1+3 graphene.}  The data in Fig.~\ref{fig:3} strongly imply the co-existence of spin domains with low and high anisotropy in t1+3.  Below, we discuss how such disparate anisotropies may naturally emerge.

In multilayer graphene, the magnetic anisotropy energy associated with collectively rotating the spin polarization of a ferromagnet depends sensitively on its valley polarization. This originates from the intrinsic Kane--Mele spin--orbit coupling, which takes the form $\sigma_z s_z \tau_z$. When the multilayer graphene lacks three-dimensional inversion symmetry~\cite{mccann2010spin}, additional spin--orbit couplings that are independent of the sublattice degree of freedom can also arise, such as terms of the form $s_z \tau_z$. Importantly, both types of spin--orbit coupling are linear in valley polarization. 
As a result, the expectation value of the spin--orbit coupling of a spin- and valley-polarized ferromagnet is finite, typically on the order of tens of \si{\micro\eV}, $\Delta_{I}\sim50$~\si{\micro\eV}~\cite{arp2024intervalley}. Such ferromagnetic states are likely the ground state at filling fractions  $\nu=\pm 3$ of moir\'e flatbands, especially when $R_{xy}\sim R_{xx}$. By contrast, a spin-polarized but valley-unpolarized ferromagnet has a vanishing first-order expectation value of the spin--orbit coupling. Its magnetic anisotropy therefore arises only at second order in the spin--orbit interaction, and is consequently much smaller. This dependence of spin anisotropy on valley polarization has been experimentally confirmed in untwisted rhombohedral multilayer graphene, where the transverse spin susceptibility of the spin-polarized, valley-unpolarized half-metal was found to be orders of magnitude larger than that of the spin-valley-polarized quarter-metal~\cite{auerbach2025visualizing}.

The spin-polarized, valley-unpolarized ferromagnet can arise in the moir\'e flat bands of twisted graphene superlattices, associated with the emergence of the $\nu=2$ insulating state, and is typically more robust than the spin-valley-polarized state. It is therefore plausible that, in the vicinity of $\nu=\pm 3$ where $\rxx$ and $\rxy$ indicate finite but not full valley polarization, domains of the spin-valley-polarized ferromagnet are interspersed with others that are spin polarized but valley unpolarized.  Then, one would naturally find adjacent domains with significantly different magnitudes of spin anisotropy.

For simplicity, we consider a situation where one domain is fully spin-valley polarized, and the other is strictly spin-polarized but valley-unpolarized. They are dubbed high-anisotropy and low-anisotropy domains, respectively. Applying a training field $\bperp$ in the positive $+\bperp$ direction initially aligns the spins in both domains, preparing the system in a low-resistance state. Because the SOC and training magnetic field favor the same spin orientation in the high-anisotropy domain, the SOC can sustain the spin order against reversal in a small opposite field $-B_{\perp}$, provided that $|g\mu_B B_{\perp}| \ll \Delta_I$. In contrast, the spins in the low-anisotropy domain flip as soon as $B$ changes sign. The resulting domain-wall configuration corresponds to a high-resistance state and leads to the MR observed in the experiment.
Furthermore, if $B_{\perp}$ is increased sufficiently to flip the spin in the high-anisotropy domain---either because the spin Zeeman energy surpasses the spin--orbit coupling energy or because the field switches the valley polarization---the spins in the two domains realign, and the system again recovers its low-resistance state. This explains why the enhanced MR is confined to a window around small negative $B$ in Fig.~\ref{fig:3}e.

\section{Acknowledgements}
The authors thank Mona Berciu, Marcel Franz, and Stuart Parkin for helpful discussions, and Silvia Folk for helpful contributions. Experiments at UBC were undertaken with support from the Natural Sciences and Engineering Research Council of Canada; the Canada Foundation for Innovation; the Canadian Institute for Advanced Research; the Max Planck-UBC-UTokyo Centre for Quantum Materials and the Canada First Research Excellence Fund, Quantum Materials and Future Technologies Program; and the European Research Council (ERC) under the European Union's Horizon 2020 research and innovation program, Grant Agreement No. 951541. Device preparation at UW was supported by National Science Foundation (NSF) CAREER award no. DMR-2041972 and NSF MRSEC 2308979. The development of twisted graphene samples was partially supported by the Department of Energy, Basic Energy Science Programs under award DE-SC0023062.
C.H is supported  by NSF CAREER grant award No.~DMR-2541471 and the Department of Energy, Office of Basic Energy Sciences under Award No.~DE-SC0024346. D.W. was supported by an appointment to the Intelligence Community Postdoctoral Research Fellowship Program at University of Washington administered by Oak Ridge Institute for Science and Education through an interagency agreement between the US Department of Energy and the Office of the Director of National Intelligence. K.W. and T.T. acknowledge support from the JSPS KAKENHI (Grant Numbers 21H05233 and 23H02052) and World Premier International Research Center Initiative (WPI), MEXT, Japan. This work made use of shared fabrication facilities at UW provided by NSF MRSEC 2308979.

\section{Author Contributions}
R.S. performed the measurements in the Folk lab at UBC, and analyzed the data; D.W. made the sample in the Yankowitz lab at UW; J.F. supervised the measurements; R.S., M.Y. and J.F. wrote the manuscript with N.W., C.H., and A.H.M. providing theory support; K.W. and T.T. provided the hBN crystals.

\section*{Competing interests}
The authors declare no competing interests.

\section*{Additional Information}
Correspondence and requests for materials should be addressed to Joshua Folk.

\section*{Data Availability}
Source data are available for this paper. All other data that support the findings of this study are available from the corresponding author upon request.

\bibliographystyle{naturemag}
\bibliography{main}
\newpage
\clearpage

\onecolumngrid

\renewcommand{\figurename}{Extended Data Fig.}
\renewcommand{\thesubsection}{S\arabic{subsection}}
\setcounter{secnumdepth}{2}
\setcounter{figure}{0} 
\setcounter{equation}{0}

\onecolumngrid

\section*{Extended Data}

\begin{figure}[ht!]
    \centering
    \includegraphics[width = \textwidth]{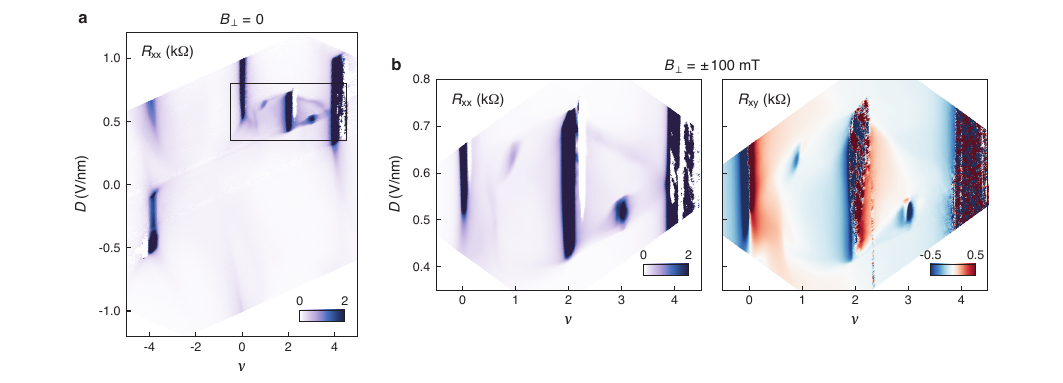}
    \caption{ \textbf{Full gate-tuned phase diagram.} \textbf{a}, $R_{xx}$ map as a function of $\nu$ and $D$ at $B_{\perp} = 0$. \textbf{b}, Symmetrized $R_{xx}$ and antisymmetized $R_{xy}$ maps at $B_{\perp} = \pm 100$ mT, focusing on $D > 0$ where the lowest \m conduction band is isolated (boxed region in \textbf{a}).}
    \label{efig:full_map_s1}
\end{figure} 

\begin{figure}[ht!]
    \centering
    \includegraphics[width = \textwidth]{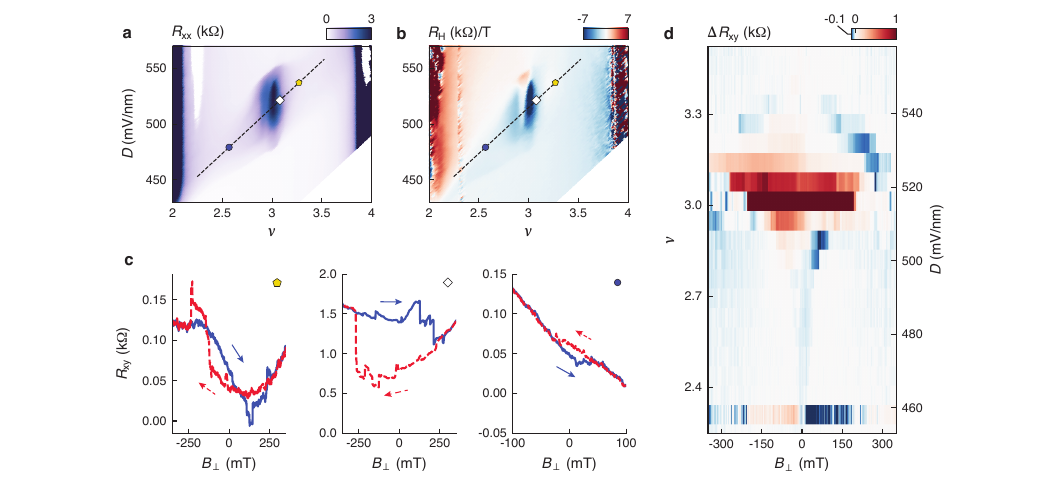}
    \caption{ \textbf{Anomalous Hall effect near $\nu = 3$.} \textbf{a}, \textbf{b}, Symmetrized $R_{xx}$ and antisymmetrized $R_{H}$ maps measured at $B_{\perp} = \pm 100$ mT, focusing on gate voltages near $\nu = 3$. Data is the same as from Fig.\ref{fig:1}b of the main text. \textbf{c}, $R_{xy}$ hysteresis loops measured at  $[\nu, D~(\mathrm{mV/nm})] = $ [3.27, 536], [3.08, 521], and [2.57, 479] corresponding to colored markers in \textbf{a,b}. \textbf{d}, Difference in $R_{xy}$ between forward and reverse magnetic field sweeps, $\Delta R_{xy} = (R_{xy}^{\uparrow} - R_{xy}^{\downarrow})$, characterizing the dependence of the AHE along a contour of fixed top gate voltage $V_{t} = 3.3$ V, and bottom gate voltages $V_{b}$ between $4.5$ and $6.0$ V, indicated by markers in \textbf{a,b}. The color scale is saturated to enhance contrast in the metallic region near $\nu = 3$.}
    \label{efig:AHE_s2}
\end{figure}

\begin{figure}[ht!]
    \centering
    \includegraphics[width = \textwidth]{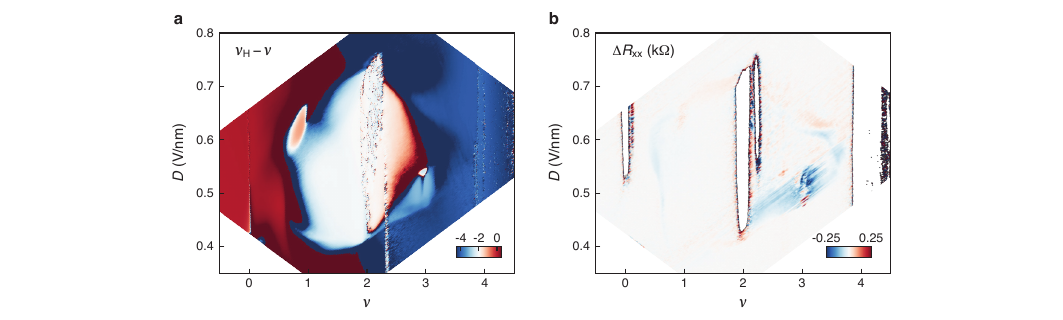}
    \caption{ \textbf{Subtracted Hall filling and negative in-plane magnetoresistance.} \textbf{a},  Map of the subtracted Hall filling, $\nu_{H} - \nu$ in the lowest \m conduction band. Data was antisymmetrized using $B_{\perp} = \pm 100$ mT. \textbf{b}, Change in the longitudinal resistance between $B_{\parallel} = 50$ mT and no applied $B_{\parallel}$, $\Delta R_{xx} = R_{xx}(B_{\parallel} = 50 \text{ mT}) - R_{xx}(0)$. Gate voltages where negative in-plane magnetoresistance is observed correspond closely to the region of isospin symmetry breaking near $\nu = 3$, where the $\nu_{H}$ (Hall filling factor) departs $\nu$ (expected \m band filling factor) by $-3$.}
    \label{efig:SHF_s3}
\end{figure}

\begin{figure}[ht!]
    \centering
    \includegraphics[width = \textwidth]{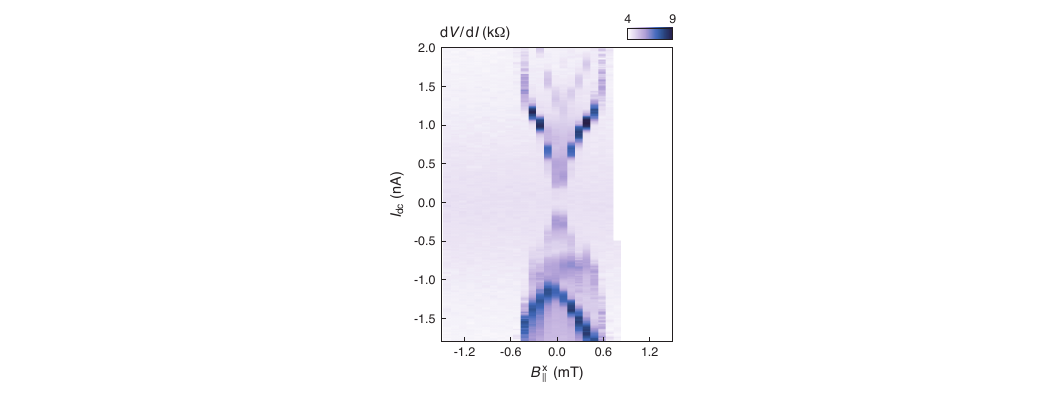}
    \caption{ \textbf{In-plane magnetic field suppression of the ferro-Josephson effect.} Current bias dependent $dV/dI$ map at $ B_{\perp} = 18$ mT, measured with $I_{ac} = 80$ pA.}
    \label{efig:suppresion_s4}
\end{figure}

\begin{figure}[ht!]
    \centering
    \includegraphics[width = \textwidth]{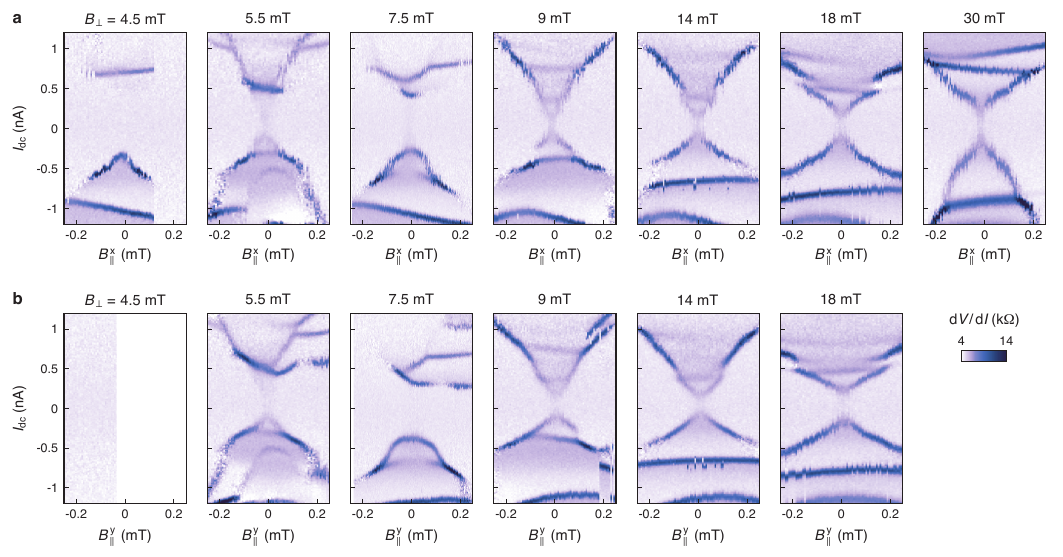}
    \caption{ \textbf{Out-of-plane magnetic field dependence of differential resistance maps.}  \textbf{a}, \textbf{b}, Bias dependent $dV/dI$ mapped out as a function of $B_{\parallel}^{x}$ and $B_{\parallel}^{y}$, at a fixed $B_{\perp}$, respectively.
    }
    \label{efig:full_dvdi}
\end{figure}

\begin{figure}[ht!]
    \centering
    \includegraphics[width = \textwidth]{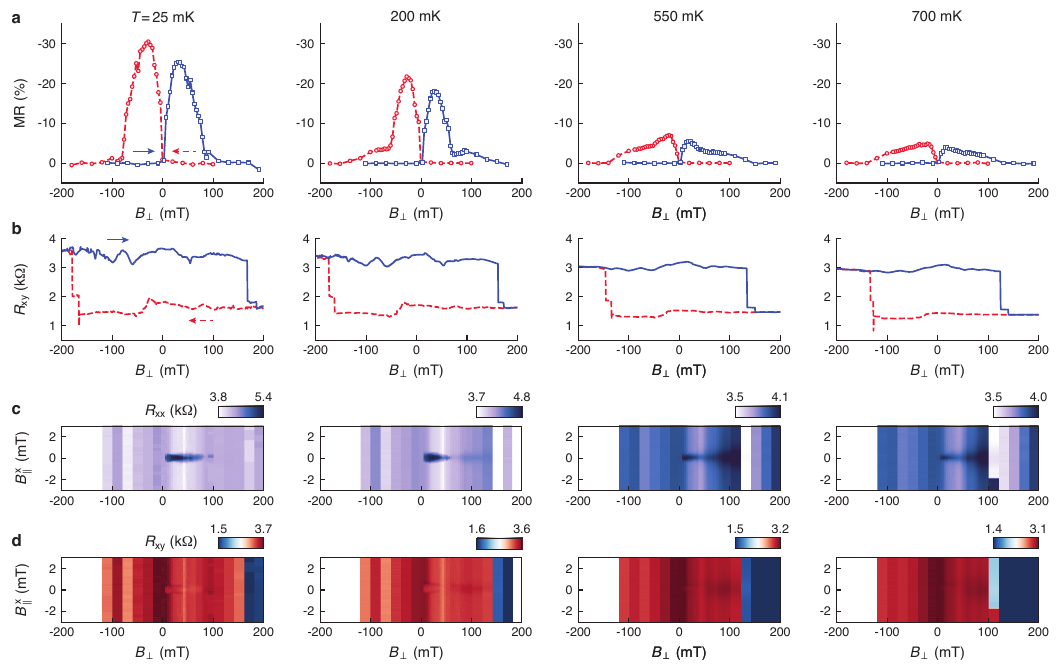}
    \caption{ \textbf{Initialization with out-of-plane magnetic field and temperature dependence.} \textbf{a, b}, In-plane MR and $R_{xy}$ as a function of $B_{\perp}$ at $\nu = 3.02, D = 0.523$ V/nm. MR was evaluated at $B_{\parallel} = 1$ mT. Arrows indicate the sweep direction of $B_{\perp}$ after ramping $\vert B_{\perp}\vert$ to 250 mT to initialize the domain structure. Data in each column (left to right) were acquired at $T = 25$, 200, 550, and 700 mK. \textbf{c}, Color maps of $R_{xx}(B_{\parallel}^{x})$ traces used to extract the MR data in \textbf{a}, recorded while reducing $B_{\perp}$ towards zero from $B_{\perp} = -250$ mT. \textbf{d}, Simultaneously recorded $R_{xy}$ corresponding to \textbf{c}.}
    \label{efig:initialization_s5}
\end{figure}

\begin{figure}[ht!]
    \centering
    \includegraphics[width = \textwidth]{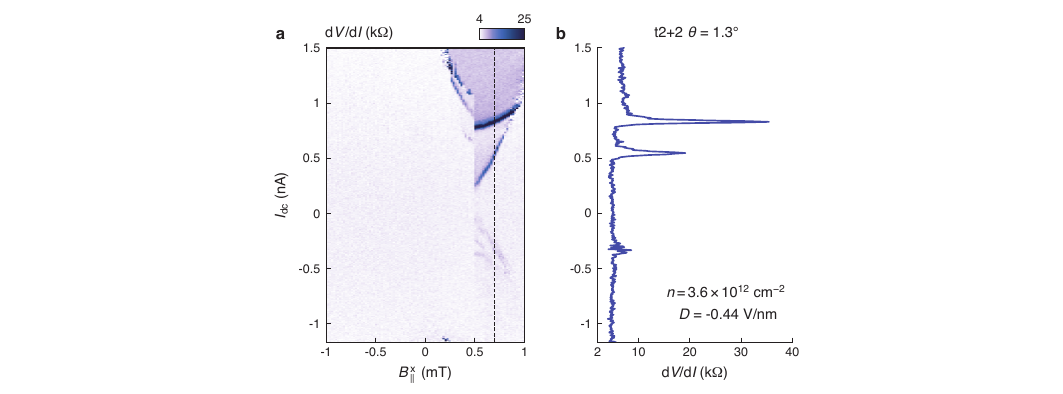}
    \caption{ \textbf{Ferro-Josephson effect in twisted double bilayer graphene.} \textbf{a}, Differential resistance map at $n = 3.6 \times 10^{12}$ cm$^{-2}$ and $D = -0.44$ V/nm in a twisted double bilayer graphene (t2+2) device. At these gate voltages, the sample exhibits the anomalous Hall effect. The same device was previously studied in Ref.\onlinecite{kuiri2022spontaneous}  (sample D2). \textbf{b} line cut at $B_{\parallel}^{x} = 0.7$ mT, showing sharp peaks in $dV/dI$ emerging with a small dc bias. These peaks respond sensitively to a weak in-plane magnetic field. Data were obtained using $I_{ac} = 10$ pA, $B_{\perp} = 0$.}
    \label{efig:tdbg_s6}
\end{figure}

\begin{figure}[ht!]
    \centering
    \includegraphics[width = \textwidth]{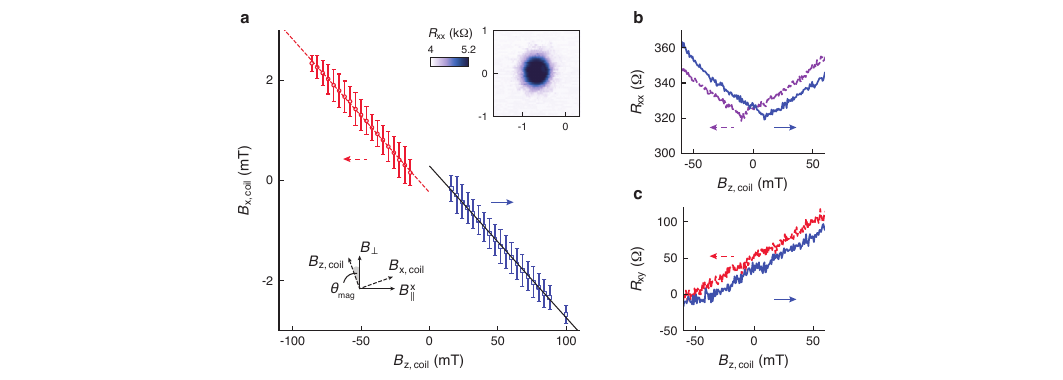}
    \caption{ \textbf{Magnetic field calibration.} \textbf{a}, Location of in-plane MR peaks at $\nu = 3.02$, $D = 0.523$ V/nm in the ($B_{\rm{x, coil}}$, $B_{\rm{z, coil}}$) plane, together with linear fits. Error bars denote the full width at half maximum of each MR peak. Data were acquired while sweeping the out-of-plane magnet axis ($B_{\rm{z, coil}}$) from $B_{\rm{z, coil}} = \pm 250$ mT; arrows indicate the sweep directions. Before field calibration, the peak positions - which we take to indicate the true zero of $B_{\parallel}^{x}$ at the device -  shift linearly in $B_{\rm{x,coil}}$ with increasing $B_{\rm{z, coil}}$. Upper inset: example of an MR peak mapped in the ($B_{\rm{x,coil}}$,$B_{\rm{y,coil}}$) plane at $B_{\rm{z,coil}} = 30$ mT. Lower inset: schematic showing the relation between the magnet coil axes and the true $B_{\perp}$ and $B_{\parallel}^{x}$ after magnetic field calibration. Together, we estimate a $\rm \theta_{\rm{mag}} = 1.75^{\circ}$ misalignment, primarily corresponding to a rotation of the magnet coil axes about the sample $B_{\parallel}^{y}$ axis. A small trapped magnetic flux associated with field sweeps is also evident from the finite $B_{\rm{x,coil}}$ intercepts of the linear fits, which we estimate to be $\pm 9.5$ mT for reverse/forward sweeps. \textbf{b}, \textbf{c}, $R_{xx}$ and $R_{xy}$ under forward and reverse $B_{\rm{z,coil}}$ sweeps at $\nu = 0.54$, $D = 0.523$ V/nm. The offset in magnetic field ($\pm 9.5$ mT) required for reverse/forward sweeps to overlap provides a second estimate of the trapped flux. Data collected $T = 25$ mK.}
    \label{efig:mag_calib}
\end{figure}

\end{document}